\documentclass[aps,prapplied,reprint,superscriptaddress,showpacs,showkeys,longbibliography]{revtex4-2}
\usepackage{amsmath,amssymb}
\usepackage{graphicx}
\usepackage{bm}
\usepackage{booktabs}
\usepackage{microtype}
\usepackage{xurl}
\usepackage[colorlinks=true,linkcolor=blue,citecolor=blue,urlcolor=blue]{hyperref}
\newcommand{\Eprint}[2]{\href{#1}{\nolinkurl{#2}}}
\graphicspath{{figures/}}
\newcommand{\ket}[1]{\lvert#1\rangle}
\newcommand{\bra}[1]{\langle#1\rvert}

\newcommand{\addrA}{Beijing National Laboratory for Condensed Matter Physics, Institute of Physics, Chinese Academy of Sciences, Beijing 100190, China}
\newcommand{\addrB}{School of Physical Sciences, University of Chinese Academy of Sciences, Beijing 100049, China}
\newcommand{\addrC}{Beijing Key Laboratory of Fault-Tolerant Quantum Computing, Beijing Academy of Quantum Information Sciences, Beijing 100193, China}
\newcommand{\addrD}{Hefei National Laboratory, Hefei 230088, China}
\newcommand{\addrE}{Beijing Key Laboratory of Advanced Quantum Technology, Beijing 100190, China}

\begin{document}
\title{Controlled-Z Gates via Floquet Coupling to Two Auxiliary States in Transmon Circuits}
\author{Yutao Chen}
\affiliation{\addrA}
\affiliation{\addrB}

\author{Jia-Chi Zhang}
\affiliation{\addrA}
\affiliation{\addrB}

\author{Tian-Ming Li}
\affiliation{\addrA}
\affiliation{\addrB}

\author{Yu Liu}
\thanks{Corresponding author: \href{mailto:yuliu@iphy.ac.cn}{yuliu@iphy.ac.cn}.}
\affiliation{\addrA}

\author{Kai Xu}
\thanks{Corresponding author: \href{mailto:kaixu@iphy.ac.cn}{kaixu@iphy.ac.cn}.}
\affiliation{\addrA}
\affiliation{\addrB}
\affiliation{\addrC}
\affiliation{\addrD}
\affiliation{\addrE}

\author{Heng Fan}
\thanks{Corresponding author: \href{mailto:hfan@iphy.ac.cn}{hfan@iphy.ac.cn}.}
\affiliation{\addrA}
\affiliation{\addrB}
\affiliation{\addrC}
\affiliation{\addrD}
\affiliation{\addrE}

\begin{abstract}
High-fidelity entangling gates are essential for quantum information processing. In transmon circuits, higher excited states provide additional pathways for engineering interactions within the computational subspace. Here we demonstrate controlled-Z (CZ) gates that exploit two noncomputational states, $\ket{02}$ and $\ket{20}$, through selective Floquet coupling. The nontrivial parametric modulation can activate both auxiliary transitions from $\ket{11}$ while suppressing the leading exchange between $\ket{01}$ and $\ket{10}$. 
Based on this approach, we develop two gate protocols:
a nonadiabatic round trip through a bright auxiliary
superposition and adiabatic evolution along a zero-energy
dark state, both achieving average gate fidelities above
$99.99\%$ in numerical simulations. 
Control of the modulation phase further extends the bright-state
protocol to conventional and unconventional geometric CZ gates.
Effective-model simulations show that unconventional control
improves tolerance to residual $ZZ$ coupling and static
qubit-frequency offsets relative to conventional control
over the parameter ranges examined.
Simulations including neighboring spectator qubits show
that the original nonadiabatic bright-state and adiabatic
dark-state CZ protocols retain high gate fidelities.
This framework offers new routes toward the experimental
realization of CZ gates and their geometric implementations.
\end{abstract}

\maketitle
\raggedbottom

\begin{figure}[!t]
\centering
\includegraphics[width=1.0\linewidth]{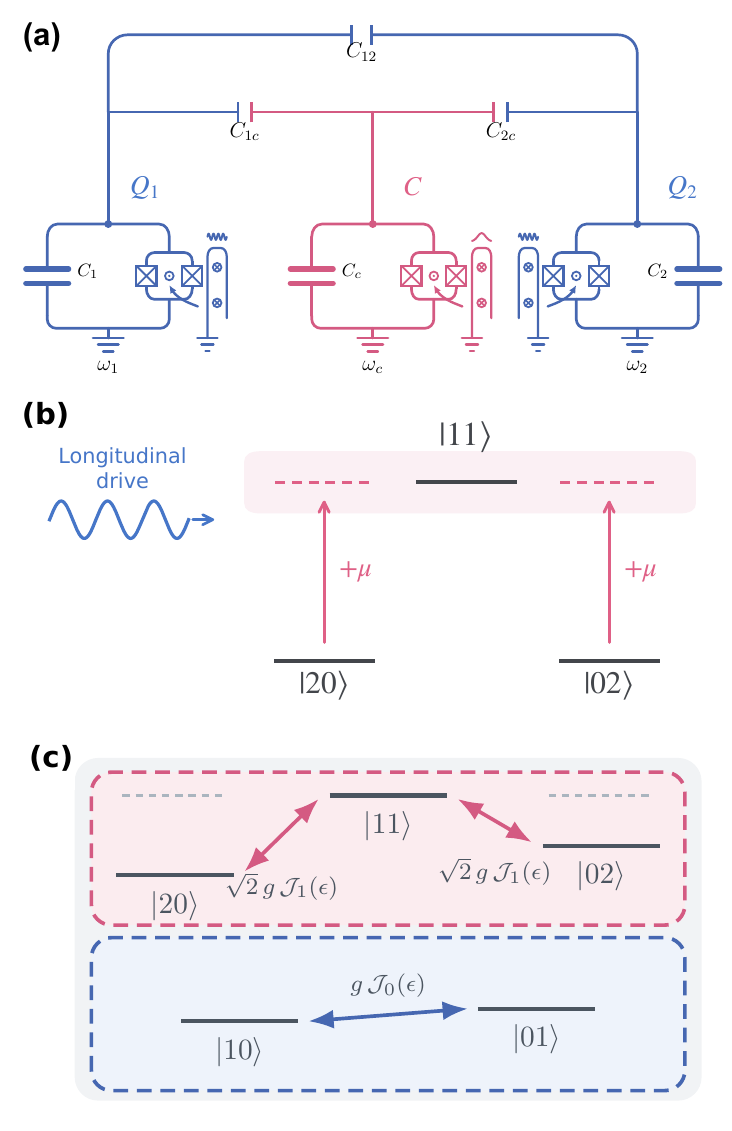}
\caption{Circuit architecture and longitudinal control for two-auxiliary-state CZ gates.
(a) Two transmons coupled directly and through a tunable coupler.
(b) Floquet resonance mechanism. At $\Delta=(\alpha_2-\alpha_1)/2$, the two bare auxiliary energies coincide relative to $\ket{11}$. In the Floquet frame, a common shift $+\mu$ brings their first-sideband replicas (dashed lines) into resonance with $\ket{11}$ when $\mu=-(\alpha_1+\alpha_2)/2$.
(c) Effective auxiliary transitions with matrix elements $\pm\sqrt{2}g\mathcal J_1(\varepsilon)$ and logical exchange proportional to $g\mathcal J_0(\varepsilon)$, where $\varepsilon=\mathcal A/\mu$. At the first zero of $\mathcal J_0$, logical exchange is suppressed to leading order while both auxiliary couplings remain finite.}
\label{fig:device}
\end{figure}

\section{Introduction}
Superconducting circuits provide a versatile platform for gate-based quantum information processing, combining engineered qubit interactions with programmable microwave and flux control~\cite{Blais2021,Krantz2019,Kjaergaard2020}. Progress in device fabrication, coherent control, and readout has enabled experiments spanning quantum algorithms~\cite{Harrigan2021,Huggins2022,RobledoMoreno2025}, digital quantum simulation~\cite{Zhang2022Floquet,Xu2024Fibonacci}, and quantum error correction~\cite{Krinner2022,Zhao2022SurfaceCode,Google2023SurfaceCode,Google2025BelowThreshold,Lacroix2025Colour,Wang2026QLDPC}. Large-scale random-circuit sampling has further demonstrated the capabilities of programmable superconducting processors~\cite{Arute2019Sycamore,Morvan2024RCS,Wu2021Zuchongzhi,Zhu2022Zuchongzhi,Gao2025Zuchongzhi}. These advances rely on programmable sequences of quantum gates. Entangling operations play a central role by establishing correlations between qubits that cannot be generated by single-qubit control alone~\cite{Barends2014}. Their physical implementation connects the available circuit interactions to the logical operations required by a quantum circuit~\cite{Foxen2020,Sung2021}. Exploring different mechanisms for generating and controlling these interactions, and understanding the conditions under which they realize an entangling gate, therefore remains an important direction in superconducting quantum control.

The controlled-Z (CZ) gate, together with single-qubit rotations, forms a universal gate set~\cite{Barends2014} and is widely used in quantum algorithms and error-correction circuits~\cite{DiCarlo2009,Google2025BelowThreshold}. A common implementation in transmons exploits the coupling of $\ket{11}$ to an auxiliary state outside the computational subspace, typically $\ket{02}$ or $\ket{20}$, to accumulate a conditional phase through adiabatic evolution or a nonadiabatic population return~\cite{DiCarlo2009,Martinis2014,Barends2019,Li2019}. 
Within this approach, 
Pulse shaping suppresses unwanted transitions and endpoint
leakage~\cite{Martinis2014,Sung2021,Rol2019,Negirneac2021},
while tunable couplers provide control over the gate coupling
and residual
interactions~\cite{Yan2018,Mundada2019,Xu2020Coupler,Chu2021,Sung2021,Li2025Calibration}.
Other implementations use resonator-mediated interactions to generate conditional phases through microwave driving~\cite{Paik2016} or continuous-variable geometric evolution~\cite{Song2017}. 
Parametric modulation  further expands the available control by selectively activating transitions~\cite{McKay2016,Caldwell2018,Reagor2018}
and enables geometric gate operations through phase
control~\cite{Chen2018a,Xu2020}, as well as superadiabatic
two-qubit gates~\cite{Chu2020}.
Holonomic entanglement generation using higher transmon levels
has also been demonstrated in an effective three-level
manifold~\cite{Egger2019}.
Together, these developments motivate exploring how selective control of noncomputational levels can be extended to the coherent participation of multiple
auxiliary states in a CZ gate.

In this work, we propose a Floquet-engineered approach to CZ gates
that coherently couples $\ket{11}$ to both auxiliary states
$\ket{02}$ and $\ket{20}$ in transmon circuits.
Periodic longitudinal modulation activates the two auxiliary
transitions while suppressing the leading exchange between
$\ket{01}$ and $\ket{10}$.
The resulting three-state manifold supports two complementary CZ gate protocols: a nonadiabatic population
cycle through a bright auxiliary superposition and adiabatic
evolution along a zero-energy dark state.
Circuit-level simulations yield coherent average gate fidelities
above $99.99\%$ for both protocols.
Control of the modulation phase further enables conventional
and unconventional geometric bright-state CZ gates.
The unconventional protocol improves tolerance
to residual $ZZ$ coupling and qubit-frequency deviations.
Finally, we examine the effects of neighboring spectator qubits
on the original bright- and dark-state protocols, motivated by
the substantial qubit-frequency excursions required during operation.
Both retain high fidelities in the configurations considered,
with a residual dependence on spectator frequencies and logical states.
Together, these results establish a unified framework for transmon CZ gates
based on selective Floquet coupling to two auxiliary states.

\section{Floquet coupling to two auxiliary states}
\label{sec:floquet}
\subsection{Circuit model and effective description}
We consider two transmon qubits, labeled $Q_1$ and $Q_2$, coupled through a controllable exchange interaction mediated by a tunable coupler, labeled $C$~\cite{Yan2018}. The system is described by the following Hamiltonian ($\hbar=1$):
\begin{align}
	\hat H_{\mathrm{QCQ}}
	={}&
	\sum_{k=1,2,c}
	\left[
	\omega_k\hat n_k
	+\frac{\alpha_k}{2}
	\hat n_k(\hat n_k-1)
	\right]\nonumber\\
	&+
	\sum_{i<j}
	g_{ij}
	(\hat a_i^\dagger-\hat a_i)
	(\hat a_j-\hat a_j^\dagger),
	\label{eq:qcq_full}
\end{align}
where $\hat a_k$ is the annihilation operator of mode
$k\in\{1,2,c\}$, where the indices $1$, $2$, and $c$ correspond to $Q_1$, $Q_2$, and $C$, respectively, and
$\hat n_k=\hat a_k^\dagger\hat a_k$. $\omega_k$ and $\alpha_k$ denote the mode
frequencies and anharmonicities, respectively, while
$g_{ij}$ characterizes the capacitive coupling between
modes $i$ and $j$, including both rotating and
counter-rotating terms. The mode frequencies can be dynamically tuned through external flux biases, $\omega_k\rightarrow\omega_k(t)$, as illustrated in Fig.~\ref{fig:device}(a).

By eliminating the coupler in the dispersive regime, we obtain an effective two-transmon Hamiltonian
\begin{align}
	\hat H_q(t)
	={}&
	\sum_{j=1,2}
	\left[
	\omega_j(t)\hat n_j
	+\frac{\alpha_j}{2}
	\hat n_j(\hat n_j-1)
	\right]\nonumber\\
	&+
	g(t)
	(\hat a_1^\dagger\hat a_2
	+\hat a_1\hat a_2^\dagger),
	\label{eq:physical}
\end{align}
where $\omega_j(t)$ and $\alpha_j$ now denote
the renormalized transmon parameters, and $g(t)$
includes both direct and coupler-mediated
exchange interactions. The relation between these effective parameters and the underlying tunable-coupler circuit is detailed in Appendix~\ref{app:circuit}.

To use both auxiliary states for a CZ gate, we seek to activate
$\ket{11}\leftrightarrow\ket{20}$ and
$\ket{11}\leftrightarrow\ket{02}$ simultaneously while suppressing
$\ket{01}\leftrightarrow\ket{10}$ exchange.
Because all three transitions arise from the same exchange term
in Eq.~\eqref{eq:physical}, controlling $g(t)$ alone cannot provide
this selectivity. Motivated by the Floquet engineering of transmon interactions demonstrated by Wang et al.~\cite{Wang2026}, we introduce longitudinal modulation
of the qubit-frequency difference,
$\omega_1(t)-\omega_2(t)=\Delta+\mathcal A\cos\mu t$,
where $\Delta$ is the mean detuning, $\mathcal A$ is the modulation
amplitude, and $\mu$ is the angular frequency~\cite{Shi2023ZeroModes,Liu2025Pumping}.
The opposite frequency excursions of the two qubits
modulate the level energies without directly driving transitions
between transmon levels. In the rotating frame, this longitudinal
drive instead gives the exchange interaction a periodic phase,
$g\exp[i\varepsilon\sin(\mu t)]$, with
$\varepsilon=\mathcal A/\mu$. Its Fourier components provide
sidebands weighted by $\mathcal J_m(\varepsilon)$, following the
frequency-modulation mechanism used to generate first-order
sidebands in circuit QED~\cite{Beaudoin2012,Strand2013}.
Choosing $\mu$ near the magnitudes of the anharmonicities brings
the first sidebands of both auxiliary transitions near resonance
[Fig.~\ref{fig:device}(b)],
while logical exchange is governed by the zeroth-order component.
This difference in sideband order allows the desired and unwanted
channels to be controlled through distinct Bessel weights.

Retaining the zeroth-order logical exchange and the first auxiliary
sidebands gives the leading-order Floquet Hamiltonian
\begin{equation}
\begin{aligned}
\hat H_F={}&\frac{\Delta}{2}
\bigl(\ket{10}\bra{10}-\ket{01}\bra{01}\bigr)\\
&+(U+\delta)\ket{20}\bra{20}
 +(U-\delta)\ket{02}\bra{02}\\
&+g\mathcal J_0(\varepsilon)
\bigl(\ket{10}\bra{01}+\mathrm{H.c.}\bigr)\\
&+\sqrt{2}g\mathcal J_1(\varepsilon)
\bigl[(\ket{20}-\ket{02})\bra{11}+\mathrm{H.c.}\bigr],
\end{aligned}
\label{eq:floquet-effective}
\end{equation}
where $\varepsilon=\mathcal A/\mu$, $\mathcal J_m$ is the Bessel
function of the first kind, $U=\mu+(\alpha_1+\alpha_2)/2$, and
$\delta=\Delta-(\alpha_2-\alpha_1)/2$.
The rotating-frame transformations and averaging conditions are
provided in Appendix~\ref{app:floquet}.
At the first Bessel zero, $\varepsilon_0\simeq2.40483$,
$\mathcal J_0(\varepsilon_0)=0$ and
$\mathcal J_1(\varepsilon_0)\simeq0.51915$, suppressing the leading
logical exchange while retaining both auxiliary couplings
[Fig.~\ref{fig:device}(c)].

\subsection{Hamiltonian in the bright--dark basis}
\label{sec:bright-dark-basis}
At $\varepsilon=\varepsilon_0$, the two auxiliary states in
Eq.~\eqref{eq:floquet-effective} can be combined into a bright
state $\ket B$ and a dark state $\ket D$~\cite{Morris1983},
\begin{equation}
 \ket B=\frac{\ket{20}-\ket{02}}{\sqrt2},\qquad
 \ket D=\frac{\ket{20}+\ket{02}}{\sqrt2}.
\end{equation}
The opposite signs of the two auxiliary couplings make their
amplitudes add for $\ket B$ and cancel for $\ket D$, so only the
bright combination couples directly to $\ket{11}$.
With $G=2g\mathcal J_1(\varepsilon_0)$, the Hamiltonian in the
ordered basis $\{\ket{11},\ket B,\ket D\}$ is
\begin{equation}
	\hat{H}_\Lambda=
	\begin{pmatrix}
		0&G&0\\
		G&U&\delta\\
		0&\delta&U
	\end{pmatrix}.
	\label{eq:bright-dark}
\end{equation}
Equation~\eqref{eq:bright-dark} provides a common description of
the two CZ gate mechanisms. At $U=\delta=0$, $\ket D$ is decoupled
and the dynamics reduce to a resonant $\ket{11}\leftrightarrow\ket B$
transition, enabling a nonadiabatic population-return cycle.
A finite $\delta$ instead couples the bright and dark auxiliary
combinations. For $U=0$, the resulting Hamiltonian supports a
zero-energy eigenstate formed from $\ket{11}$ and $\ket D$;
adiabatically varying $G$ and $\delta$ provides the second CZ route.
We develop the pulse conditions and gate dynamics for these two
protocols below.

\begin{figure*}[t]
	\centering
	\includegraphics[width=0.9\linewidth]{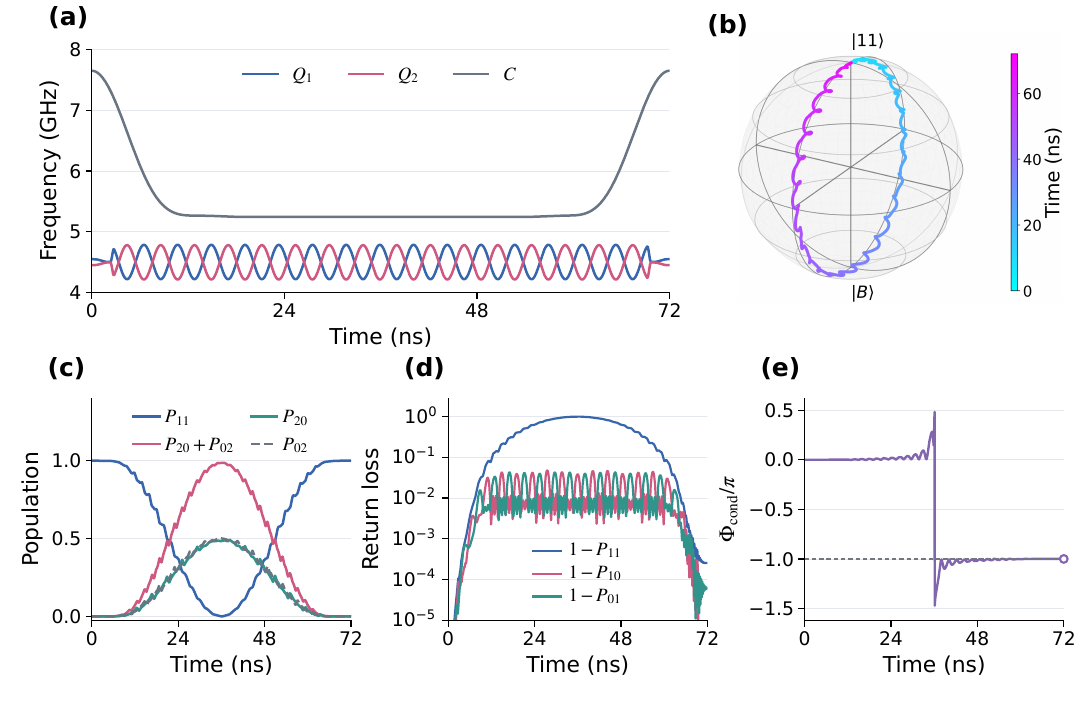}
	\caption{Nonadiabatic bright-state CZ gate in the full circuit model.
		(a) Frequency waveforms of qubits $Q_1$, $Q_2$, and coupler $C$.
		(b)  Bloch-sphere trajectory illustrating the bright-state population-transfer-and-return cycle.
		(c) Dressed-state populations showing transfer from $\ket{\widetilde{11}}$ to both auxiliary states and subsequent return. 
		(d) Leakage and swap errors, quantified by the return losses $1-P_{11}$ and $1-P_{ij}$ ($ij=10,01$), respectively.
		(e) Conditional-phase evolution toward $\pi$ modulo $2\pi$.
		The $72\,\mathrm{ns}$ gate achieves an average fidelity exceeding $99.99\%$.}
	\label{fig:bright-route}
\end{figure*}

\section{CZ protocol by nonadiabatic bright-state evolution}
\label{sec:bright}

\subsection{Resonant evolution and conditional phase}

We first consider the resonant bright-state protocol by setting
$U=\delta=0$ in Eq.~\eqref{eq:bright-dark}, corresponding to the resonance
conditions $\mu=-(\alpha_1+\alpha_2)/2$ and
$\Delta=(\alpha_2-\alpha_1)/2$. The dynamics reduce to the resonantly
coupled $\{\ket{11},\ket B\}$ subspace, while the dark state
$\ket D$ remains decoupled.
Starting from $\ket{11}$, the time evolution is given by
\begin{equation}
	\begin{aligned}
		\ket{\psi(t)}
		&=\cos\Theta(t)\ket{11}
		-i\sin\Theta(t)\ket B,\\
		\Theta(t)
		&=\int_0^t G(t')\,dt'.
	\end{aligned}
\end{equation}
The condition $\Theta(T)=\pi$ completes a coherent
population-return cycle, yielding
$\ket{11}\rightarrow-\ket{11}$
with no residual auxiliary-state occupation in the ideal model.
For a constant coupling, the ideal return time is
$T=\pi/|G|$, with $G=2g\mathcal{J}_1(\varepsilon_0)$. At the leading-order Bessel-zero operating point,
exchange between $\ket{10}$ and $\ket{01}$ is suppressed.
Together with the phase of $\pi$ acquired by $\ket{11}$ during
the bright-state cycle, this yields a CZ gate up to local phases,
which can be compensated by virtual-$Z$ rotations~\cite{McKay2017VirtualZ,Sung2021}.

\subsection{Numerical implementation and results}
\label{sec:numerical}
\label{sec:numerics}
To assess the bright-state protocol beyond the effective two-state
approximation, we simulate the complete gate evolution under the full Hamiltonian in Eq.~\eqref{eq:qcq_full}, including the finite-duration
ramps between the idle and working points.
Each transmon and the coupler is truncated to its four lowest levels,
yielding a Hilbert-space dimension of $4^3=64$.
The controls are parameterized by the mode-frequency trajectories,
which can be mapped to flux pulses through the device-specific
flux--frequency relations.
We use a parameter set with idle frequencies \(\omega_{1,\mathrm{idle}}/2\pi=4.55\,\mathrm{GHz}\), \(\omega_{2,\mathrm{idle}}/2\pi=4.45\,\mathrm{GHz}\), and \(\omega_{c,\mathrm{idle}}/2\pi=7.65\,\mathrm{GHz}\). The anharmonicities are \(\alpha_1/2\pi=-250\,\mathrm{MHz}\), \(\alpha_2/2\pi=-240\,\mathrm{MHz}\), and \(\alpha_c/2\pi=-250\,\mathrm{MHz}\). The coupling strengths are \(g_{12}/2\pi=4\,\mathrm{MHz}\) and \(g_{1c}/2\pi=g_{2c}/2\pi=100\,\mathrm{MHz}\). 
 
 The bare-qubit detuning \(\Delta_q(t)/2\pi\) is ramped from
 \(100\,\mathrm{MHz}\) at idle to \(5\,\mathrm{MHz}\) at the working
 point and back using error-function ramps~\cite{Yuan2026}, and we set the parameter
 \(\sigma_q\) as the characteristic time scale.
  During the interaction window, the two qubit frequencies are
  sinusoidally modulated under a smooth envelope.
  Meanwhile, the coupler frequency follows a compressed-Slepian
  trajectory from idle to its working point and back, with edge
  shaping inspired by Ref.~\cite{Martinis2014}.
  These frequency waveforms are shown in
  Fig.~\ref{fig:bright-route}(a), the details are given in
  Appendix~\ref{app:waveforms}.

To examine the bright-state dynamics in the full physical system,
we focus on the evolution of the qubit 
$\{\ket{11},\ket B\}$ subspace introduced
in the effective description, as illustrated by the
Bloch-sphere representation in Fig.~\ref{fig:bright-route}(b).
The trajectory illustrates a nearly complete
population-transfer-and-return cycle between
$\ket{11}$ and the Floquet bright state
$\ket B$, consistent with the resonant two-state dynamics.
The Bloch trajectory is plotted in a rotating frame,
with residual micromotion retained. The frame transformations
and projection conventions are specified in
Appendix~\ref{app:floquet}.
Figure~\ref{fig:bright-route}(c) shows the evolution of the
dressed-state populations.
With the system initialized in the idle dressed state
$\ket{\widetilde{11}}$, predominantly corresponding to
the two-qubit configuration $\ket{11}$, population is
transferred into both auxiliary states
$\ket{\widetilde{20}}$ and $\ket{\widetilde{02}}$
before returning to the initial state.
The simultaneous participation of the two auxiliary
channels supports the bright-state mechanism predicted
by the effective Floquet Hamiltonian.
Together, these results demonstrate that the
bright-state population-return dynamics remain dominant
in the full QCQ evolution.

We next verify that this mechanism realizes a CZ gate and
 evaluate its performance on the entire computational
subspace. Specifically, 
we quantify
the agreement between the numerical operation $U$ and the
ideal CZ operation $U_{\mathrm{CZ}}$ using the
average gate fidelity\cite{PEDERSEN200747}:
\begin{equation}
	\bar F =
	\frac{
		\left|\operatorname{Tr}(U_{\mathrm{CZ}}^\dagger U)\right|^2
		+\operatorname{Tr}(U^\dagger U)
	}{d(d+1)},
	\label{eq:avg_fidelity}
\end{equation}
where $d = 2^n$ for an n-qubit system.  
By optimizing the gate parameters, we evaluate
the leakage and swap errors, defined as
$\varepsilon_{\mathrm{leak}}=1-P_{11}$
and $\varepsilon_{\mathrm{swap}}=1-P_{ij}$
($ij=10,01$), respectively, where
$P_{ij}$ denotes the return probability
of the corresponding initial dressed state.
As shown in Fig.~\ref{fig:bright-route}(d),
the large transient population transfer from
$\ket{\widetilde{11}}$ is followed by a nearly
complete return, while the single-excitation states
maintain high return probabilities.
Beyond the low leakage and swap error, Fig.~\ref{fig:bright-route}(e)
shows that the conditional phase approaches $\pi$
at the end of the gate.
These results demonstrate a CZ gate with a duration
of $72\,\mathrm{ns}$ and an average fidelity
exceeding $99.99\%$.

\begin{figure*}[t]
	\centering
	\includegraphics[width=0.85\linewidth]{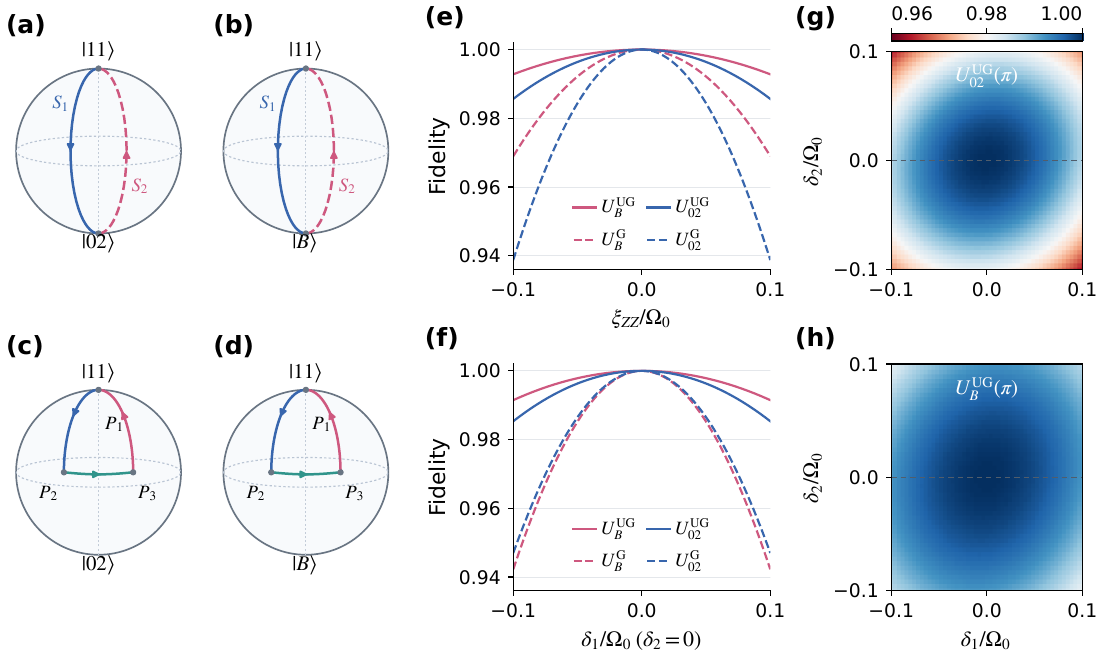}
	\caption{Bloch-sphere trajectories and robustness of conventional and unconventional geometric $\pi$ gates.
		(a,c) Trajectories in $\{|11\rangle,|02\rangle\}$; (b,d) trajectories in $\{|11\rangle,|B\rangle\}$, where $|B\rangle=(|20\rangle-|02\rangle)/\sqrt{2}$.
		Conventional trajectories (a,b) follow $S_1$ and $S_2$, while unconventional trajectories (c,d) follow $P_1\rightarrow P_2\rightarrow P_3\rightarrow P_1$.
		(e,f) Fidelity versus $\xi_{ZZ}/\Omega_0$ and $\delta_1/\Omega_0$ ($\delta_2=0$), respectively. Pink/blue curves represent the $|B\rangle$/$|02\rangle$ channels; solid/dashed lines denote unconventional/conventional gates.
		(g,h) Fidelity maps of $U_{02}^{\mathrm{UG}}(\pi)$ and $U_B^{\mathrm{UG}}(\pi)$ versus $\delta_1/\Omega_0$ and $\delta_2/\Omega_0$, with a shared color scale and $\Omega_0=\Omega_{02}$. Dashed lines mark $\delta_2=0$, corresponding to the UG curves in (f).}
	\label{fig:geometric-robustness}
\end{figure*}

\subsection{Geometric control of the bright-state CZ gate}
\label{sec:geometric_cz}
The resonant protocol above realizes a CZ gate through a
population-return cycle with a fixed coupling phase.
To explore whether additional control of the evolution path can
improve robustness, we now allow the phase of the longitudinal
modulation to vary, extending the same bright-state interaction
to geometric CZ protocols~\cite{Chen2018a,Xu2020,Chen2018}.
Replacing $\cos\mu t$ by $\cos[\mu t+\varphi(t)]$ preserves the
relative sign of the two auxiliary couplings while changing their
common phase. At $U=\delta=0$, the leading-order Hamiltonian becomes
\begin{equation}
\hat H_B(t)=\frac{\Omega_B(t)}{2}
\left[e^{-i\varphi(t)}\ket{11}\bra B+\mathrm{H.c.}\right],
\label{eq:bright_geometric}
\end{equation}
where $\Omega_B(t)=4g(t)\mathcal J_1(\varepsilon_0)$.
Thus, $\varphi(t)$ controls the rotation axis in the
$\{\ket{11},\ket B\}$ subspace. For comparison, the single-channel
protocol replaces $\ket B$ by $\ket{02}$ and uses
$\Omega_{02}(t)=2\sqrt{2}g(t)\mathcal J_1(\varepsilon_0)$.
Both couplings retain the Bessel renormalization.

A conventional geometric gate uses two resonant segments, each
with pulse area $\int\Omega(t)\,dt=\pi$.
For equal coupling phases, the state completes the closed path
$S_1\rightarrow S_2$ in Fig.~\ref{fig:geometric-robustness}(a,b),
returning to $\ket{11}$ with geometric phase $\pi$ and zero
dynamical phase. We denote the resulting gates by
$U_{02}^{\mathrm G}(\pi)$ and $U_B^{\mathrm G}(\pi)$.
The unconventional protocol instead follows the three-segment
path $P_1\rightarrow P_2\rightarrow P_3\rightarrow P_1$
in Fig.~\ref{fig:geometric-robustness}(c,d)~\cite{Chen2026}.
Here the dynamical and geometric phases satisfy
$\gamma_d=\eta\gamma_g$, following the unconventional geometric
phase relation~\cite{Zhu2003}, so that the total phase
$\gamma=(1+\eta)\gamma_g=\pi$ yields
$U_{02}^{\mathrm{UG}}(\pi)$ or $U_B^{\mathrm{UG}}(\pi)$.
The rotating-frame construction and segment conditions are given
in Appendix~\ref{app:geometric_derivation}.

We compare these four protocols in the six-dimensional effective
Hilbert space at the same single-channel coupling
$\Omega_0=\Omega_{02}$, retaining either one or both auxiliary
transitions. For the unconventional gates, we use
$\chi=0.43\pi$ and $\gamma=\pi$ following Ref.~\cite{Chen2026}.
Residual $ZZ$ coupling is modeled by
\begin{equation}
\hat H_{ZZ}=\frac{\xi_{ZZ}}{2}
\left(\ket{00}\bra{00}-\ket{01}\bra{01}
-\ket{10}\bra{10}+\ket{11}\bra{11}\right).
\end{equation}
Figure~\ref{fig:geometric-robustness}(e) shows that the bright-state
gates are more robust against this perturbation than their
single-channel counterparts, and that unconventional control
further improves the fidelity for both configurations.

\begin{figure*}[t]
	\centering
	\includegraphics[width=0.9\linewidth]{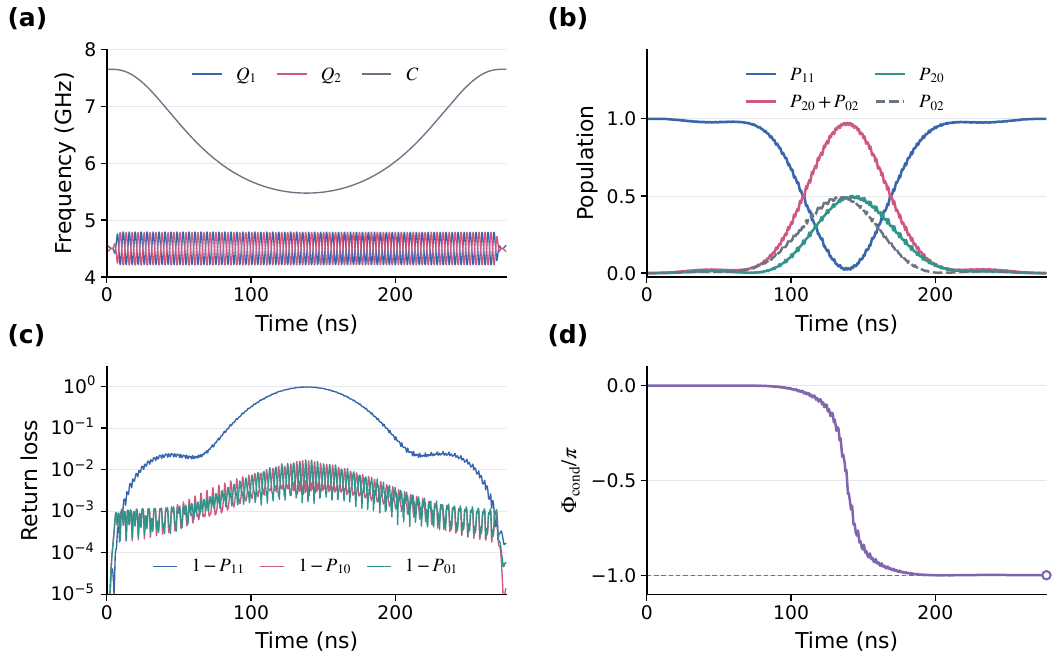}
	\caption{Adiabatic dark-state CZ gate in the full circuit model. (a) Frequency waveforms of qubits $Q_1$, $Q_2$, and coupler $C$. (b) Populations projected onto the fixed idle-dressed states during evolution from $|\widetilde{11}\rangle$. (c) Logarithmic return losses, $1-P_{11}$, $1-P_{10}$, and $1-P_{01}$, for the corresponding logical inputs. (d) conditional phase $\Phi_{\mathrm{cond}}/\pi$; the dashed line marks the CZ target at $-1$, and the circle marks the final phase. The total pulse duration is $277\,\mathrm{ns}$, and the average gate fidelity is exceeding $99.99\%$.}
	\label{fig:smooth-idle-cz}
\end{figure*}

For qubit-frequency deviations, the comparison depends on the
control protocol and detuning direction.
Along $\delta_2=0$ [Fig.~\ref{fig:geometric-robustness}(f)],
the conventional bright-state gate is slightly less robust than
the single-channel gate: differential detuning couples the bright
and dark states and opens an additional leakage channel.
Unconventional control suppresses the final dark-state population
and improves the fidelity, with the bright-state protocol
outperforming its single-channel counterpart in the scans shown
in Fig.~\ref{fig:geometric-robustness}(f--h).
The conventional gates' two-dimensional comparison and the
residual dark-state populations are discussed in
Appendix~\ref{app:conventional_detuning}.
These results identify phase control as a means of improving
bright-state CZ robustness within the effective model.
The static-error scans considered here do not characterize the
full response to time-dependent flux noise. That response also
depends on the modulation operating point and noise spectrum,
as established in studies of ac sweet spots and parametric-gate
noise sensitivity~\cite{Didier2019AC,Hong2020AC,Ganzhorn2020}.

\begin{figure*}[t]
	\centering
	\includegraphics[width=\textwidth]{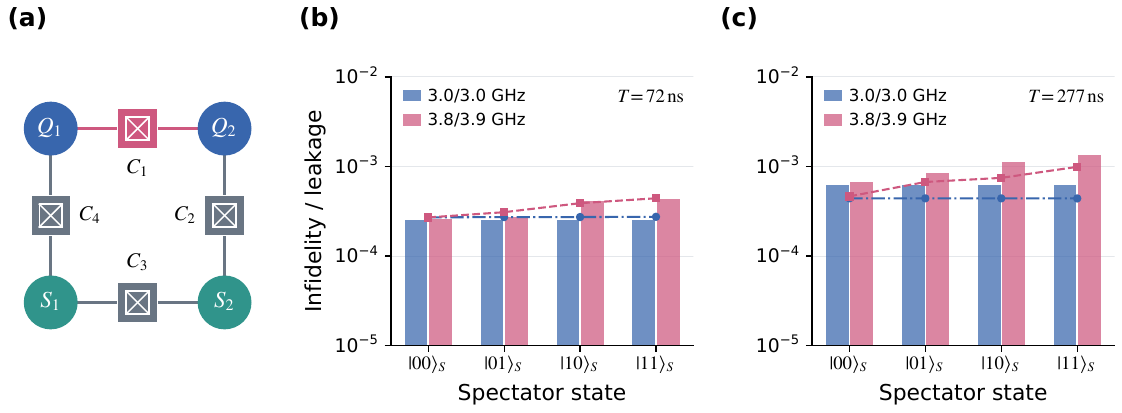}
	\caption{Spectator effects on bright- and dark-state CZ gates.
		\textbf{(a)} Simulated circuit with active qubits $Q_1,Q_2$,
		spectator qubits $S_1,S_2$, and couplers $C_1$--$C_4$.
		The active coupling path through $C_1$ is highlighted in pink.
		\textbf{(b)} Nonadiabatic bright-state CZ gate with
		$T=72\,\mathrm{ns}$.
		\textbf{(c)} Adiabatic dark-state CZ gate with total duration
		$T=277\,\mathrm{ns}$.
		Bars show the gate infidelity $1-F$.
		Lines show the leakage error, averaged over
		the four active logical inputs.
		The horizontal labels denote spectator logical states.
		Blue and pink colors correspond to spectator-frequency
		pairs $(3.0,3.0)\,\mathrm{GHz}$ and
		$(3.8,3.9)\,\mathrm{GHz}$, respectively.}
	\label{fig:spectator_cz}
\end{figure*}

\section{CZ protocol by adiabatic dark-state evolution}
\label{sec:dark}
\subsection{Dark-state path and conditional phase}
In contrast to the nonadiabatic bright-state protocol, which relies on a resonant cycle, the same three-state manifold spanned by $\ket{11}$, $\ket{20}$, and $\ket{02}$ can also be configured as an effective $\Lambda$ system in the ${\ket{11},\ket B,\ket D}$ basis, enabling STIRAP-like adiabatic passage analogous to that in atomic and trapped-ion systems~\cite{Duan2001, Pachos2004}. Building on this analogy and inspired by studies of STIRAP geometric control~\cite{Setiawan2023} and doublon-mediated geometric gates in optical lattices~\cite{Kiefer2026}, We now develop an adiabatic dark-state protocol for realizing a CZ gate, exploiting the detuning $\delta(t)$ and the coupling $G(t)$ to control the instantaneous dark eigenstate. Setting $U=0$, the effective Hamiltonian in $\{\ket{11},\ket B,\ket D\}$ reads:
\begin{equation}
	H_{\mathrm{ad}}(t)=
	\begin{pmatrix}
		0 & G(t) & 0 \\
		G(t) & 0 & \delta(t) \\
		0 & \delta(t) & 0
	\end{pmatrix}.
	\label{eq:adiabatic_hamiltonian}
\end{equation}
 the Hamiltonian admits a zero-energy dark eigenstate for arbitrary $G(t)$ and $\delta(t)$, provided they do not vanish simultaneously:
 \begin{equation}
 	\ket{D_{\mathrm{ad}}(t)}
 	=\frac{\delta(t)\ket{11}-G(t)\ket D}
 	{\sqrt{G^2(t)+\delta^2(t)}}.
 	\label{eq}
 \end{equation}
Here the subscript $\mathrm{ad}$ distinguishes the instantaneous dark eigenstate from the auxiliary dark state $\ket D$. This eigenstate has no $\ket B$ component due to destructive interference between the two coupling pathways.
Introducing the mixing angle $\theta(t)$ through $\tan\theta(t)=G(t)/\delta(t)$, with a continuous choice of branch, the dark state can be written as $\ket{D_{\mathrm{ad}}(t)}=\cos\theta(t)\ket{11}-\sin\theta(t)\ket D$.
By varying $\theta(t)$ from $0$ to $\pi$ under the adiabatic condition $|\dot{\theta}(t)|\ll\Omega(t)$, where $\Omega(t)=\sqrt{G^2(t)+\delta^2(t)}$ is the instantaneous energy gap, the system follows the dark eigenstate and evolves from $\ket{11}$ to $-\ket{11}$. 
In the ideal effective frame, the dark-state eigenenergy remains zero throughout the evolution, so no dynamical phase is accumulated. The sign change upon completing the adiabatic path gives rise to a geometric phase of $\pi$. 
Provided that the $\ket{01}\leftrightarrow\ket{10}$ exchange remains suppressed during the Floquet modulation, the geometric phase acquired by $\ket{11}$ produces the desired conditional phase of $\pi$ in the effective model. After compensating the remaining single-qubit phases by local $Z$ rotations, the resulting operation is equivalent to $U_{\mathrm{CZ}}=\mathrm{diag}(1,1,1,-1)$. To assess the feasibility of this protocol beyond the ideal effective description, we next numerically examine the dynamics and conditional phase in the full physical model.

\subsection{Numerical implementation and results}

Following the approach adopted for the bright-state protocol in Sec.~\ref{sec:numerical}, we implement the adiabatic dark-state protocol with the same circuit parameters. The corresponding
waveforms are shown in Fig.~\ref{fig:smooth-idle-cz}(a).
During the interaction window, the detuning control $\delta(t)$
follows a cosine profile whose argument varies smoothly from
$0$ to $\pi$ according to a beta cumulative distribution
function, taking the detuning between equal and opposite
endpoint values.
The auxiliary coupling $G(t)$ is controlled through the
coupler-frequency waveform, which is determined using the
calibrated relation between the effective qubit exchange
coupling and the coupler frequency.
Further details are given in 
Appendix~\ref{app:darkwaveforms}.

The population dynamics shown in Fig.~\ref{fig:smooth-idle-cz}(b)
can be understood from the dark-state composition
in Eq.~\eqref{eq}.
In the ideal description, the dark state is dominated by
$\ket{11}$ when $|G(t)|\ll|\delta(t)|$, whereas near
$\delta(t)=0$ its weight shifts to
$\ket D=(\ket{20}+\ket{02})/\sqrt{2}$.
The comparable populations of the two auxiliary states near
the middle of the numerical pulse are consistent with this
structure, with their combined occupation reaching
approximately $0.974$.
As the detuning changes sign and the coupling is reduced,
the ideal dark-state path returns to $-\ket{11}$.
The numerical evolution correspondingly shows a near-complete
recovery of $\ket{\widetilde{11}}$ with a small residual error, while keeping the final swap errors below $10^{-4}$
[Fig.~\ref{fig:smooth-idle-cz}(c)]. This behavior is consistent with the Floquet modulation
suppressing logical exchange while retaining the auxiliary
couplings needed for the gate. The accompanying accumulation of a conditional phase close to
$\pi$ [Fig.~\ref{fig:smooth-idle-cz}(d)]
completes the intended CZ action and the resulting gate achieves an average fidelity of $99.9941\%$ with a total duration
of $277\,\mathrm{ns}$.

\section{Spectator effects on bright- and dark-state CZ gates}
\label{sec:spectator}
The CZ-gate implementations in Secs.~\ref{sec:bright}
and~\ref{sec:dark} involve qubit-frequency modulation
with amplitudes of order $10^2\,\mathrm{MHz}$.
This frequency-tuning requirement presents a practical
limitation in multiqubit circuits, where the active qubits
may approach transitions of neighboring qubits or parasitic
two-level systems and experience unwanted interactions.
To assess spectator-induced errors in this setting,
we examine the two calibrated gates in a four-qubit square 
lattice with tunable couplers~\cite{Yuan2026},
as illustrated in Fig.~\ref{fig:spectator_cz}(a).

The CZ gate acts on the active qubits $Q_1,Q_2$,
while the spectator qubits $S_1,S_2$ and the additional
couplers $C_2$--$C_4$ remain at their idle points.
Each protocol retains its previously calibrated active
parameters and control waveforms.
For each spectator-frequency configuration,
we evaluate all four spectator logical states.
The endpoint leakage $L$ denotes the population outside
the computational subspace of $Q_1,Q_2$,
averaged over the four active logical inputs.

When both spectators are at $3.0\,\mathrm{GHz}$,
the gate error and leakage show little variation
with the spectator state
[Figs.~\ref{fig:spectator_cz}(b,c)].
The $72\,\mathrm{ns}$ bright-state gate maintains
an average fidelity $F>99.95\%$, while the
$277\,\mathrm{ns}$ dark-state gate achieves $F>99.9\%$.
For both protocols, the leakage remains below
$5\times10^{-4}$ in every spectator state.
Thus, the gate performance is only weakly dependent
on the neighboring logical states in this configuration.

Raising the spectator frequencies to
$(f_{S_1},f_{S_2})=(3.8,3.9)\,\mathrm{GHz}$
produces a more pronounced state dependence,
particularly for the dark-state gate.
The largest gate error and leakage occur for
$\ket{11}_S$ in both protocols.
Nevertheless, the bright-state gate maintains
$F>99.95\%$ and $L<5\times10^{-4}$ for all
spectator states, while the dark-state gate retains
$F>99.8\%$ and $L<10^{-3}$.
With the active control waveforms unchanged,
this variation demonstrates that residual spectator
interactions depend on the surrounding frequency
configuration. The frequencies and logical states
of neighboring qubits therefore remain relevant
to gate calibration in multiqubit implementations.

\section{Conclusion}
We have presented a Floquet-engineered approach to CZ gates
that uses both auxiliary states $\ket{20}$ and $\ket{02}$
in transmons with negative and unequal anharmonicities.
Longitudinal modulation selectively activates their couplings
to $\ket{11}$ while suppressing the leading single-excitation
exchange. The resulting bright--dark basis provides a common
description of two complementary gate mechanisms: a resonant
population-return cycle through the bright state and adiabatic
evolution along a zero-energy dark eigenstate.
Circuit-model simulations yield coherent average gate fidelities
above $99.99\%$ after local-$Z$ corrections, with durations of
$72\,\mathrm{ns}$ and $277\,\mathrm{ns}$ for the bright- and
dark-state protocols, respectively.

Control of the modulation phase also enables conventional and
unconventional geometric implementations of the bright-state CZ gate.
Effective-model comparisons show that the unconventional protocol
improves tolerance to residual $ZZ$ coupling and suppresses
detuning-induced population remaining in the dark state.
The response to qubit-frequency errors nevertheless depends on
the control path and detuning direction, so robustness is not
guaranteed by the participation of both auxiliary transitions alone.
These results identify the evolution path as an additional control
resource within the same Floquet-engineered manifold.

With the active gate waveforms fixed, simulations including
neighboring spectator qubits show that gate performance depends
on both spectator frequencies and logical states.
This dependence connects the isolated-pair protocols to the
frequency-allocation and calibration requirements of multiqubit
circuits. Together, our results establish coherent control of
two auxiliary transitions as a framework for realizing CZ gates
through distinct population-transfer and geometric evolution paths.

\par
\begingroup
\vspace*{12pt}
\makeatletter
\def\section{\@startsection{section}{1}{\z@}{0pt}{8pt}{\normalfont\small\bfseries\centering}}
\makeatother
\begin{acknowledgments}
We appreciate the support from the Synergetic Extreme
Condition User Facility (SECUF) in Huairou District, Beijing.
This work was supported by the National Natural Science
Foundation of China (Grants No. U25A6009, No. 92265207, No. 92565301, No., 92365301,
No. T2121001, No. T2322030, No. 12122504, No. 12274142, No. 12475017, and No. 12604773), the Innovation Program for Quantum
Science and Technology (Grant No. 2021ZD0301800), the
Beijing Nova Program (Grant No. 20220484121), the
Ministry of Science and Technology project (Grant No.
2025YFE0217600), and the China Postdoctoral Science Foundation (Grants No. GZC20252227, 2025M783453, and 2026T190913).
\end{acknowledgments}
\endgroup

\bibliography{main}

\clearpage
\appendix

\section{Circuit model and dispersive reduction}
\label{app:circuit}

Here we detail the dispersive reduction connecting
Eqs.~\eqref{eq:qcq_full} and~\eqref{eq:physical}.
For convenience, we define
$\Delta_{jc}=\omega_j-\omega_c$ and
$\Sigma_{jc}=\omega_j+\omega_c$.
In the dispersive regime $|g_{jc}|,|\alpha_\mu|\ll|\Delta_{jc}|$,
we assume that the coupler remains in its ground state and apply
the Schrieffer--Wolff transformation
$\widetilde H=e^{\hat S}\hat H_{\mathrm{QCQ}}e^{-\hat S}$
with~\cite{Yan2018}
\begin{equation}
	\hat S=
	\sum_{j=1,2}
	\left[
	\frac{g_{jc}}{\Delta_{jc}}
	\hat a_j^\dagger\hat a_c
	-\frac{g_{jc}}{\Sigma_{jc}}
	\hat a_j^\dagger\hat a_c^\dagger
	-\mathrm{H.c.}
	\right].
\end{equation}
To second order in the qubit--coupler couplings, the
effective Hamiltonian takes the static form of
Eq.~\eqref{eq:physical} after the rapidly rotating pair terms
are neglected. The effective exchange coupling is
\begin{equation}
	g\simeq g_{12}
	+\frac{g_{1c}g_{2c}}{2}
	\sum_{j=1,2}
	\left(
	\frac{1}{\Delta_{jc}}
	-\frac{1}{\Sigma_{jc}}
	\right).
	\label{eq:dispersive-exchange}
\end{equation}
The local dispersive frequency shifts are absorbed into
$\omega_j$, following the notation in the main text,
while corrections to $\alpha_j$ are neglected at this order.
The $1/\Sigma_{jc}$ terms account for virtual
counter-rotating processes.

External fluxes tune the Josephson energies and hence the
mode frequencies, $\omega_\mu\rightarrow\omega_\mu(t)$
\cite{Koch2007,Krantz2019}.
For approximately constant $g_{ij}$, the instantaneous
exchange becomes
\begin{equation}
	g(t)\simeq g_{12}
	+\frac{g_{1c}g_{2c}}{2}
	\sum_{j=1,2}
	\left[
	\frac{1}{\Delta_{jc}(t)}
	-\frac{1}{\Sigma_{jc}(t)}
	\right],
	\label{eq:dispersive-exchange-time}
\end{equation}
where
$\Delta_{jc}(t)=\omega_j(t)-\omega_c(t)$ and
$\Sigma_{jc}(t)=\omega_j(t)+\omega_c(t)$
are evaluated from the circuit frequencies in
Eq.~\eqref{eq:qcq_full}.
The corresponding local frequency shifts are absorbed into
$\omega_j(t)$ in Eq.~\eqref{eq:physical}.
This instantaneous approximation requires dispersive separation
throughout the pulse and negligible time-derivative corrections,
with $|\dot\Delta_{jc}(t)|\ll|\Delta_{jc}(t)|^2$
for constant $g_{jc}$~\cite{Roth2017}.

\section{Effective Floquet Hamiltonian and Bloch-vector representation}
\label{app:floquet}

\subsection{Derivation of the effective Hamiltonian}

The exchange interaction in Eq.~\eqref{eq:physical} drives both
logical exchange and the auxiliary transitions, so varying its
strength alone does not select the desired channels.
For negative transmon anharmonicities, the bare auxiliary
resonance conditions, $\Delta+\alpha_1=0$ and
$-\Delta+\alpha_2=0$, cannot be satisfied simultaneously~\cite{Yuan2026}.
Longitudinal modulation addresses both constraints: its periodic
phase on the exchange interaction generates Floquet sidebands,
allowing the auxiliary transitions to be selected independently
of the leading logical exchange.

In the working region, take the effective exchange $g$ in Eq.~\eqref{eq:physical} to be locally constant and write the longitudinal modulation as
\begin{equation}
 \omega_{1,2}(t)=\bar\omega
 \pm\frac{\Delta}{2}\pm\frac{\mathcal A}{2}\cos\mu t,
 \qquad \varepsilon=\frac{\mathcal A}{\mu}.
 \label{eq:floquet-longitudinal-drive}
\end{equation}
Here $\bar\omega$ is the mean effective transmon frequency. With $\hat N_q=\hat n_1+\hat n_2$ and $\hat D_q=\hat n_1-\hat n_2$, the common-frequency and longitudinal-drive rotations combine into
\begin{equation}
 \hat V_1(t)=e^{-i[\bar\omega t\hat N_q+f(t)\hat D_q]},\qquad
 f(t)=\frac{\mathcal A}{2\mu}\sin\mu t.
 \label{eq:floquet-local-frame}
\end{equation}
For $\ket{\psi_1}=\hat V_1^\dagger\ket{\psi_q}$, the transformed Hamiltonian is
\begin{equation}
 \begin{aligned}
  \hat H_1(t)={}&\hat V_1^\dagger\hat H_q(t)\hat V_1
  -i\hat V_1^\dagger\dot{\hat V}_1\\
  ={}&\frac{\Delta}{2}\hat D_q
  +\sum_{j=1,2}\frac{\alpha_j}{2}\hat n_j(\hat n_j-1)\\
  &+g\left[e^{2if(t)}\hat a_1^\dagger\hat a_2
  +e^{-2if(t)}\hat a_1\hat a_2^\dagger\right].
 \end{aligned}
 \label{eq:floquet-modulated-exchange}
\end{equation}
The derivative term cancels the common-frequency and longitudinal-modulation terms. The Jacobi--Anger expansion gives
\begin{equation}
 e^{\pm i\varepsilon\sin\mu t}
 =\sum_{m=-\infty}^{\infty}
 \mathcal J_m(\pm\varepsilon)e^{im\mu t}.
 \label{eq:sideband-expansion}
\end{equation}

We next introduce the auxiliary-state rotation
\begin{equation}
 \hat V_2(t)=e^{i\mu t\hat P_a},\qquad
 \hat P_a=\ket{20}\bra{20}+\ket{02}\bra{02}.
 \label{eq:floquet-sideband-frame}
\end{equation}
With transformed Hamiltonian,
\begin{equation}
 \hat H_2(t)=\hat V_2^\dagger\hat H_1(t)\hat V_2+\mu\hat P_a,
 \label{eq:floquet-shifted-hamiltonian}
\end{equation}
where the last term shifts both auxiliary-state energies by $\mu$. Averaging over one modulation period, $\langle F\rangle_\mu=(\mu/2\pi)\int_0^{2\pi/\mu}F(t)\,dt$, yields
\begin{equation}
 \begin{aligned}
  \overline{H}_{10,01}
  &=g\langle e^{i\varepsilon\sin\mu t}\rangle_\mu
   =g\mathcal J_0(\varepsilon),\\
  \overline{H}_{20,11}
  &=\sqrt2g\langle e^{i\varepsilon\sin\mu t-i\mu t}\rangle_\mu
   =\sqrt2g\mathcal J_1(\varepsilon),\\
  \overline{H}_{02,11}
  &=\sqrt2g\langle e^{-i\varepsilon\sin\mu t-i\mu t}\rangle_\mu
   =-\sqrt2g\mathcal J_1(\varepsilon).
 \end{aligned}
 \label{eq:channels}
\end{equation}
The auxiliary matrix elements select $m=1$ in Eq.~\eqref{eq:sideband-expansion}, while their Hermitian conjugates select $m=-1$. Their opposite signs follow from $\mathcal J_1(-\varepsilon)=-\mathcal J_1(\varepsilon)$ in this frame convention.

Together with the diagonal energies $U\pm\delta$, these period-averaged terms give Eq.~\eqref{eq:floquet-effective}. The simultaneous-resonance condition $U=\delta=0$ requires
\begin{equation}
 \mu_{\rm res}=-\frac{\alpha_1+\alpha_2}{2},\qquad
 \Delta_{\rm res}=\frac{\alpha_2-\alpha_1}{2}.
 \label{eq:resonance-app}
\end{equation}
At $\varepsilon=\varepsilon_0$, the leading logical exchange vanishes. In the two-excitation manifold, transforming to the bright--dark basis $\{\ket{11},\ket B,\ket D\}$ gives Eq.~\eqref{eq:bright-dark}, with $G=2g\mathcal J_1(\varepsilon_0)$. This leading-order averaging requires the exchange, residual auxiliary detunings, and envelope variation rates to be small compared with $\mu$, with the discarded harmonics off resonance.

The opposite signs of the auxiliary matrix elements make their
contributions add for $\ket B=(\ket{20}-\ket{02})/\sqrt2$ and
cancel for $\ket D=(\ket{20}+\ket{02})/\sqrt2$.
At $\delta=0$, $\ket D$ is decoupled; setting $U=0$ also makes
$\ket{11}\leftrightarrow\ket B$ resonant.
A finite $\delta$ mixes the bright and dark auxiliary combinations,
producing the chain $\ket{11}\leftrightarrow\ket B\leftrightarrow\ket D$.
For $U=0$, destructive interference yields a zero-energy eigenstate
formed from $\ket{11}$ and $\ket D$, with no bright-state component.
Varying $G$ and $\delta$ adiabatically therefore enables dark-state
passage analogous to STIRAP~\cite{Bergmann1998,Kis2002,Moller2007,Vitanov2017,Xu2016STIRAP}.

\subsection{Dressed-state basis and projection frame}
The full-QCQ state $\ket{\psi_{\rm lab}(t)}$ is initialized in the idle dressed state $\ket{\widetilde{11}}$.
The reference states satisfy $\hat H_{\rm idle}\ket{\widetilde{jk}}=E_{jk}^{\rm idle}\ket{\widetilde{jk}}$, with $jk\in\{11,20,02\}$ and the coupler label suppressed.
Figure~\ref{fig:bright-route}(c) shows the laboratory-frame populations $P_{\widetilde{jk}}=|\langle\widetilde{jk}|\psi_{\rm lab}\rangle|^2$.
In Fig.~\ref{fig:bright-route}(b), the label $\ket B$ denotes the fixed ideal antisymmetric projection state
\begin{equation}
 \ket{\widetilde B_-}=\frac{\ket{\widetilde{20}}-\ket{\widetilde{02}}}{\sqrt2}.
 \label{eq:ideal-dressed-bright-state}
\end{equation}

To specify the projection frame, write the longitudinal drive in Eq.~\eqref{eq:qcq-waveforms} as $A w(t)\cos\chi(t)\hat D_q$, where $\hat D_q=\hat n_1-\hat n_2$ and $\chi(t)=\nu(t-r_q)$.
Here $\nu$ is an angular frequency and $A$ is the single-qubit amplitude, with $\mathcal A=2A$ in Eq.~\eqref{eq:floquet-longitudinal-drive}.
The full-pulse primitive and its canonical form are
\begin{equation}
 \eta(t)=A\int_0^t w(s)\cos\chi(s)\,ds,\qquad
 \eta_{\rm can}(t)=\eta(t)-C.
 \label{eq:qcq-drive-primitive}
\end{equation}
For any plateau time $t_p$, choose $C=\eta(t_p)-(A/\nu)\sin\chi(t_p)$.
This single constant gives $\eta_{\rm can}=(A/\nu)\sin\chi$ on the plateau while retaining the envelope integral on both edges.

Define the longitudinal rotation $\hat W(t)=e^{i\eta_{\rm can}(t)\hat D_q}$ and the fixed dressed auxiliary projector $\widetilde P_a=\ket{\widetilde{20}}\bra{\widetilde{20}}+\ket{\widetilde{02}}\bra{\widetilde{02}}$.
The auxiliary rotation $\hat Q(t)=e^{-i\chi(t)\widetilde P_a}$ subtracts the prescribed carrier phase $\chi(t)$ from both auxiliary states, yielding a clearer Bloch trajectory.
The full state used for projection is
\begin{equation}
 \ket{\psi_P(t)}=\hat Q(t)\hat W(t)\ket{\psi_{\rm lab}(t)}.
 \label{eq:qcq-projection-frame}
\end{equation}
Its Hamiltonian, including both frame-derivative terms, is
\begin{equation}
 \hat H_P=
 \hat Q\left[\hat W\hat H_{\rm QCQ}\hat W^\dagger
 -\dot\eta_{\rm can}\hat D_q\right]\hat Q^\dagger
 +\nu\widetilde P_a.
 \label{eq:qcq-sideband-projection-hamiltonian}
\end{equation}
Here $\dot\eta_{\rm can}=A w\cos\chi$ cancels the explicit longitudinal drive.
The projector $\widetilde P_a$ uses idle dressed states, unlike
$\hat P_a$ in Eq.~\eqref{eq:floquet-sideband-frame}.
These rotations are fixed by the waveform and reference basis and are applied to the full state before projection.

\subsection{Projected Bloch coordinates}
Let $a=\langle\widetilde{11}|\psi_P\rangle$ and $b=\langle\widetilde B_-|\psi_P\rangle$, with retained weight $N=|a|^2+|b|^2$.
In the ordered basis $\{\ket{\widetilde{11}},\ket{\widetilde B_-}\}$, the unnormalized Bloch coordinates are
\begin{equation}
 X=2\operatorname{Re}(a^*b),\qquad
 Y=2\operatorname{Im}(a^*b),\qquad
 Z=|a|^2-|b|^2.
 \label{eq:projected-bloch-coordinates}
\end{equation}
If $a_{\rm dr}$ and $b_{\rm dr}$ denote the projections before $\hat Q$, then $a=a_{\rm dr}$ and $b=e^{-i\chi}b_{\rm dr}$.
Thus the auxiliary rotation changes the azimuth by the known carrier phase while preserving $Z$ and $N$.
For this projected pure state, $\sqrt{X^2+Y^2+Z^2}=N$, the radius records the weight retained in the chosen two-state projection.
As shown in Fig.~\ref{fig:bright-route}(b), this resulting trajectory is consistent with the resonant dynamics predicted by the effective bright-state model.

\section{Control waveforms for the CZ controls}
\label{app:waveforms}
\subsection{Bright-state proctocol}
This appendix specifies the control waveforms used to implement the bright-state and adiabatic dark-state CZ protocols. For each protocol, we describe the control trajectories, pulse envelopes, and timing conventions used in the numerical simulations.
The bright-state CZ gate in
Sec.~\ref{sec:numerics}, is illustrated in
Fig.~\ref{fig:bright-route}(a). The complete pulse takes the
system from the idle configuration to the working region and
back, combining a qubit-detuning sweep, opposite-sign
longitudinal modulation of the two qubits, and a shaped coupler
trajectory. For a total pulse duration \(T\), the three mode
frequencies are parameterized as
\begin{equation}
	\begin{aligned}
		\omega_{q_1}(t)
		&=\bar{\omega}_q+\frac{\Delta_q(t)}{2}
		+A w(t)\cos\!\left[\nu(t-r_q)\right],\\
		\omega_{q_2}(t)
		&=\bar{\omega}_q-\frac{\Delta_q(t)}{2}
		-A w(t)\cos\!\left[\nu(t-r_q)\right],\\
		\omega_c(t)
		&=\omega_{c,\mathrm{idle}}
		+\bigl(\omega_{c,\mathrm{work}}-\omega_{c,\mathrm{idle}}\bigr)
		s_{\mathrm{cS}}(t).
	\end{aligned}
	\label{eq:qcq-waveforms}
\end{equation}
Here \(\bar{\omega}_q\) is the fixed mean qubit frequency,
\(\Delta_q(t)\) is the bare-qubit detuning before adding the
modulation, and \(A\) and \(\nu\) are the single-qubit modulation
amplitude and angular frequency, respectively.

The detuning \(\Delta_q(t)/(2\pi)\) is ramped from
\(100\,\mathrm{MHz}\) at idle to \(5\,\mathrm{MHz}\) at the
working point and back using error-function ramps~\cite{Yuan2026},
with characteristic time scale \(\sigma_q\).
The modulation envelope \(w(t)\) is nonzero only within
\(r_q<t<T-r_q\), where \(r_q=4\sqrt{2}\,\sigma_q\).
Its turn-on and turn-off edges are constructed from the
quintic smootherstep function
\(S_5(x)=10x^3-15x^4+6x^5\).
The time \(r_q\) also defines the phase origin of the modulation.

The coupler frequency is controlled independently through the
normalized trajectory \(s_{\mathrm{cS}}(t)\), which joins the
idle and working frequencies with symmetric edges and a central
plateau. We use a compressed-Slepian parameterization with
finite-cosine edges inspired by Ref.~\cite{Martinis2014}:
 {\small
\begin{equation}
	\begin{aligned}
		s_{\mathrm{cS}}(t)
		&=
		\begin{cases}
			s_{\mathrm{Sl}}\!\left(\dfrac{t}{rT};c_2,c_3\right),
			& 0\le t<\dfrac{rT}{2},\\[4pt]
			1,
			& \dfrac{rT}{2}\le t\le T-\dfrac{rT}{2},\\[4pt]
			s_{\mathrm{Sl}}\!\left(\dfrac{T-t}{rT};c_2,c_3\right),
			& T-\dfrac{rT}{2}<t\le T,
		\end{cases}\\
		s_{\mathrm{Sl}}(x;c_2,c_3)
		&=
		\frac{\displaystyle\sum_{k=1}^{3}c_k[1-\cos(2\pi kx)]}
		{\displaystyle\sum_{k=1}^{3}c_k[1-\cos(\pi k)]},
		\qquad c_1=1 .
	\end{aligned}
\end{equation}
}
The parameter \(r\) sets the duration \(rT/2\) of each coupler
edge, while \(c_2\) and \(c_3\) determine its shape, with
\(c_1=1\).
\subsection{Adiabatic dark-state proctocol}
\label{app:darkwaveforms}
We assume the interaction onset as $t=0$ and denote the
interaction duration by $T_{\mathrm a}$.
To control $\delta(t)$, we sweep the slow bare-qubit
detuning about $\Delta_0=(\alpha_2-\alpha_1)/2$, choosing
$\Delta_q(t)=\Delta_0+r\cos\vartheta(s)$, where
$s=t/T_{\mathrm a}\in[0,1]$.
The sweep angle is parameterized as
\begin{equation}
	\begin{aligned}
		\vartheta(s)&=\pi I_s(\beta,\beta),\\
		I_s(\beta,\beta)
		&=
		\frac{\displaystyle\int_0^s
			u^{\beta-1}(1-u)^{\beta-1}\,du}
		{\displaystyle\int_0^1
			u^{\beta-1}(1-u)^{\beta-1}\,du}.
	\end{aligned}
	\label{eq:dark-detuning-schedule}
\end{equation}
Here, $I_s(\beta,\beta)$ is the regularized incomplete beta
function, and $\beta>1$ adjusts the sweep rate while ensuring
smooth endpoints.
As $\vartheta$ increases from $0$ to $\pi$, the detuning
offset changes from $+r$ to $-r$.

Equation~\eqref{eq:dispersive-exchange} shows how the exchange
coupling can be controlled through the coupler frequency.
For numerical waveform construction, we calibrate the effective coupling $g_{\mathrm{cal}}(\omega_c,\Delta_q)$ from the static 
Hamiltonian and determine $\omega_c(t)$
by solving
\begin{equation}
	g_{\mathrm{cal}}\!\left(\omega_c(t),\Delta_q(t)\right)
	=
	g_{\mathrm{idle}}\!\left[\Delta_q(t)\right]
	-g_{\mathrm{exc}}[\sin\vartheta(s)]^p,
	\label{eq:dark-inverse-calibration}
\end{equation}
where
$g_{\mathrm{idle}}(\Delta_q)
=g_{\mathrm{cal}}(\omega_{c,\mathrm{idle}},\Delta_q)$
denotes the small residual coupling at the idle point for the current slow detuning.
The constant $g_{\mathrm{exc}}>0$ sets the maximum exchange
excursion, while $p$ controls its shape.
At both endpoints, the excursion vanishes, ensuring that
the coupler is at the idle frequency.
Near the middle of the interaction window, the coupling
strength reaches its maximum, with the coupler at its lowest
frequency and the detuning offset passing through zero.

The loading and reset ramps connect the slow detuning
to its idle value using the $S_5$ function defined in the
preceding subsection, with durations $t_{\mathrm{load}}$
and $t_{\mathrm{reset}}$, respectively.
The same function shapes the modulation envelope $w(t)$:
it rises from zero to unity and falls back to zero through
edges of duration $\tau_w$, with $w(t)=1$ between the edges.
During loading and reset, the longitudinal modulation is
off and the coupler remains at its idle frequency.

The opposite longitudinal modulations applied to the qubits
are $\pm Aw(t)\cos(\nu t)$, where $A$ is the single-qubit
modulation amplitude and $\nu$ is the angular frequency.
The corresponding auxiliary coupling is approximately
\begin{equation}
	G(t)\simeq
	2g_{\mathrm{cal}}\!\left(\omega_c(t),\Delta_q(t)\right)
	\mathcal J_1\!\left[\frac{2Aw(t)}{\nu}\right].
	\label{eq:dark-effective-coupling}
\end{equation}

\begin{figure*}[t]
	\centering
	\includegraphics[width=\textwidth]
	{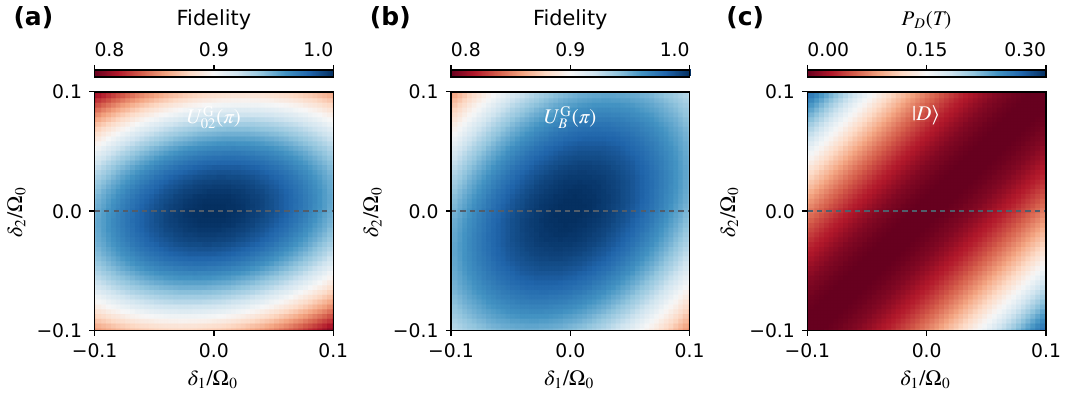}
	\caption{
		Frequency-detuning response
		of conventional geometric CZ gates.
		Average gate fidelities of
		(a) the single-channel gate and
		(b) the bright-state gate.
		(c) Residual dark-state population $P_D(T)$
		of the bright-state gate for an initial
		$\ket{11}$ state.
		Dashed lines indicate $\delta_2=0$,
		corresponding to the one-dimensional scans
		in Fig.~\ref{fig:geometric-robustness}(f).
	}
	\label{fig:conventional_detuning}
\end{figure*}

\section{Unconventional geometric gate}
\subsection{Derivation of the unconventional geometric control conditions}
\label{app:geometric_derivation}

For a linearly varying coupling phase $\varphi(t)=at+\varphi_0$,
we use the transformation
$R_{\Delta_e}(t)=e^{-i\Delta_e t/2}\ket{11}\bra{11}
+e^{i\Delta_e t/2}\ket B\bra B$.
The transformed phase is $\varphi_e(t)=(a-\Delta_e)t+\varphi_0$.
The complete three-segment control conditions are
\begin{widetext}
\begin{equation}
	\begin{aligned}
		{}[0,\tau_1):\quad&
		\Omega_B\tau_1=\chi,\quad
		\varphi_e=\frac{\pi}{2},\quad \Delta_e=0\\
		{}[\tau_1,\tau_2):\quad&
		\Omega_B(\tau_2-\tau_1)
		=\frac{2\gamma\sin\chi\,d_\chi}
		{(1+\eta)(1-\cos\chi)},\quad \varphi_e(t)=
		-\frac{\Omega_B(t-\tau_1)}
		{\sin\chi\,d_\chi},\\
		&\Delta_e=\Omega_B
		\left(\frac{1}{\sin\chi\,d_\chi}
		-\cot\chi\right),\\
		{}[\tau_2,\tau):\quad&
		\Omega_B(\tau-\tau_2)=\chi,\quad \varphi_e=
		-\frac{2\gamma}{(1+\eta)(1-\cos\chi)}
		-\frac{\pi}{2},\quad \Delta_e=0,
	\end{aligned}
	\label{eq:unconventional_geometric_pulse}
\end{equation}
\end{widetext}
where $d_\chi=(1+\eta)\cos\chi-\eta$.

We derive the three-segment control conditions for
the unconventional nonadiabatic geometric CZ gate
in the effective $\{\ket{11},\ket B\}$ subspace.
The effective Hamiltonian is
\begin{equation}
	\frac{H_{\mathrm{eff}}(t)}{\hbar}
	=
	-\frac{\Delta_e(t)}{2}\sigma_z^{(B)}
	+
	\frac{\Omega_B}{2}
	\left(
	e^{-i\varphi_e(t)}\ket{11}\bra B
	+\mathrm{H.c.}
	\right),
	\label{eq:app_geo_hamiltonian}
\end{equation}
where
$\sigma_z^{(B)}=\ket{11}\bra{11}-\ket B\bra B$.
We assume a constant Rabi frequency $\Omega_B$
within the three control segments.
Starting from $\ket{11}$, the evolving state
can be parameterized as
\begin{equation}
	\ket{\psi(t)}
	=
	e^{if(t)}
	\left[
	\cos\frac{\chi(t)}{2}\ket{11}
	+
	e^{i\xi(t)}
	\sin\frac{\chi(t)}{2}\ket B
	\right],
	\label{eq:app_geo_state}
\end{equation}
where $\chi(t)$ and $\xi(t)$ denote the polar
and azimuthal angles on the Bloch sphere,
respectively.
Substituting Eq.~\eqref{eq:app_geo_state}
into the Schrödinger equation gives
\begin{subequations}
	\begin{align}
		\dot{\chi}(t)
		&=
		\Omega_B\sin[\varphi_e(t)-\xi(t)],
		\label{eq:app_chi_dot}\\
		\dot{\xi}(t)
		&=
		-\Delta_e(t)
		-\Omega_B\cot\chi(t)
		\cos[\varphi_e(t)-\xi(t)].
		\label{eq:app_xi_dot}
	\end{align}
\end{subequations}

We construct a closed trajectory consisting of
three segments. In the first segment $P_1\rightarrow P_2$,
we choose $\Delta_e=0$ and
$\varphi_e=\xi_1+\pi/2$.
Equations~\eqref{eq:app_chi_dot} and
\eqref{eq:app_xi_dot} then give
$\dot{\chi}=\Omega_B$ and $\dot{\xi}=0$.
Thus, the state reaches a polar angle $\chi$
at $t=\tau_1$, satisfying $\Omega_B\tau_1=\chi$.

In the second segment $P_2\rightarrow P_3$, the state evolves
along a latitude with fixed polar angle $\chi$.
Setting $\varphi_e(t)=\xi(t)$ ensures
$\dot{\chi}=0$, while
Eq.~\eqref{eq:app_xi_dot} becomes
$	\dot{\xi}
	=
	-\Delta_e-\Omega_B\cot\chi.
$
Writing $\dot{\xi}=-\beta$, we obtain
\begin{equation}
	\Delta_e=\beta-\Omega_B\cot\chi,
	\qquad
	\varphi_e(t)=\xi_1-\beta(t-\tau_1).
	\label{eq:app_latitude_control}
\end{equation}
The total phase acquired during the cyclic
evolution consists of geometric and
dynamical contributions$
	\gamma=\gamma_g+\gamma_d.
$
The geometric phase is determined by the
solid angle enclosed by the trajectory,
\begin{equation}
	\gamma_g
	=
	-\frac{1}{2}
	\oint(1-\cos\chi)\,d\xi
	=
	\frac{\beta(\tau_2-\tau_1)}{2}
	(1-\cos\chi).
	\label{eq:app_gamma_g}
\end{equation}
Since the dynamical phase vanishes along
the two meridional segments, its contribution
comes entirely from $P_2$:
\begin{equation}
	\begin{aligned}
		\gamma_d
		&=
		-\frac{1}{\hbar}
		\int_{\tau_1}^{\tau_2}
		\bra{\psi(t)}H_{\mathrm{eff}}(t)
		\ket{\psi(t)}\,dt\\
		&=
		-\frac{\tau_2-\tau_1}{2}
		\left(
		\Omega_B\sin\chi-\Delta_e\cos\chi
		\right)\\
		&=
		\frac{\tau_2-\tau_1}{2}
		\left(
		\beta\cos\chi-\frac{\Omega_B}{\sin\chi}
		\right).
	\end{aligned}
	\label{eq:app_gamma_d}
\end{equation}
For an unconventional geometric gate, we impose
the phase relation
$
	\gamma_d=\eta\gamma_g.
$
Combining Eqs.~\eqref{eq:app_gamma_g}--
\eqref{eq:app_gamma_d}, we find
\begin{equation}
	\beta=
	\frac{\Omega_B}{\sin\chi\,d_\chi},
	\qquad
	d_\chi=(1+\eta)\cos\chi-\eta.
	\label{eq:app_beta}
\end{equation}
Consequently, the control parameters in $P_2$
satisfy
\begin{equation}
	\begin{aligned}
		\varphi_e(t)
		&=
		\xi_1-
		\frac{\Omega_B(t-\tau_1)}
		{\sin\chi\,d_\chi},\\
		\Delta_e
		&=
		\Omega_B
		\left(
		\frac{1}{\sin\chi\,d_\chi}
		-\cot\chi
		\right).
	\end{aligned}
	\label{eq:app_middle_control}
\end{equation}
Using $\gamma=(1+\eta)\gamma_g$, the duration
of the second segment is determined by
\begin{equation}
	\Omega_B(\tau_2-\tau_1)
	=
	\frac{2\gamma\sin\chi\,d_\chi}
	{(1+\eta)(1-\cos\chi)}.
	\label{eq:app_middle_duration}
\end{equation}
The final azimuthal angle is therefore
\begin{equation}
	\xi_2
	=
	\xi_1-
	\frac{2\gamma}
	{(1+\eta)(1-\cos\chi)}.
	\label{eq:app_xi2}
\end{equation}

In the final segment $P_3\rightarrow P_1$, we set
$\Delta_e=0$ and
$\varphi_e=\xi_2-\pi/2$.
The resulting evolution satisfies
$\dot{\chi}=-\Omega_B$ and $\dot{\xi}=0$,
returning the state to $\ket{11}$ after
$
	\Omega_B(\tau-\tau_2)=\chi.
$

With the phase reference $\xi_1=0$,
these results reproduce the three-segment
control conditions in
Eq.~\eqref{eq:unconventional_geometric_pulse}.
The cyclic evolution gives
\begin{equation}
	\ket{11}
	\longrightarrow
	e^{i(\gamma_g+\gamma_d)}\ket{11}.
\end{equation}
Setting $\gamma=\pi$ yields the required
conditional phase for the effective-frame
geometric CZ gate.
\subsection{Two-dimensional detuning robustness of
	conventional geometric gates}
\label{app:conventional_detuning}

To complement the one-dimensional detuning analysis
in the main text, Fig.~\ref{fig:conventional_detuning}(a,b)
presents the two-dimensional fidelity landscapes of
the conventional geometric gates.
Although the bright-state gate exhibits slightly
lower fidelity along $\delta_2=0$, it outperforms
the single-channel gate over most of the scanned
parameter region.

In the $\{\ket{11},\ket{B},\ket{D}\}$ subspace,
the frequency-deviation term becomes
\begin{equation}
	\frac{H_{\delta}}{\hbar}
	=
	(\delta_1+\delta_2)I_3
	+(\delta_1-\delta_2)
	(\ket{B}\bra{D}+\ket{D}\bra{B}),
	\label{eq:app_bright_dark_detuning}
\end{equation}
revealing the bright-dark coupling induced by
differential detuning.
Figure~\ref{fig:conventional_detuning}(c)
shows the corresponding residual dark-state
population,
$P_D(T)=|\bra{D}U(T)\ket{11}|^2$,
which vanishes along $\delta_1=\delta_2$
and increases with differential detuning.
Its maximum reaches approximately $27.2\%$
for the conventional gate, compared with
$1.57\%$ for the unconventional gate under
the same detuning range.
This comparison indicates that the unconventional
protocol substantially suppresses the residual
dark-state population at the end of the gate.
\end{document}